\documentclass{iopart}

\usepackage{graphicx}
\usepackage{subfig}
\usepackage{caption}
\usepackage{upgreek}
\usepackage{color}
\usepackage{amsfonts}
\usepackage{amssymb}
\usepackage{gensymb}
\usepackage{setspace}
\newcommand{\ket}[1]{\left|{#1}\right>}
\newcommand{\bra}[1]{\left<{#1}\right|}

\begin{document}
\title[Verification of state and entanglement with incomplete tomography]{Verification of state and entanglement with incomplete tomography}
\author{Yong~Siah~Teo~$^1$~$^2$, Berthold-Georg~Englert~$^1$~$^3$,
Jaroslav~{\v R}eh{\'a}{\v c}ek~$^4$, Zden{\v e}k~Hradil~$^4$, and Dmitry~Mogilevtsev~$^5$}
\address{$^1$ Centre for Quantum Technologies, National University of Singapore, Singapore 117543, Singapore}
\address{$^2$ NUS Graduate School for Integrative Sciences and Engineering, Singapore 117597, Singapore}
\address{$^3$ Department of Physics, National University of Singapore, Singapore 117542, Singapore}
\address{$^4$ Department of Optics, Palack{\'y} University, 17. listopadu 12, 77146 Olomouc, Czech Republic}
\address{$^5$ Institute of Physics, Belarus National Academy of Sciences, F. Skarina Ave. 68, Minsk 220072, Belarus}
\ead{hradil@optics.upol.cz}
\begin{abstract}
There exists, in general, a convex set of quantum state estimators that maximize the likelihood for informationally incomplete data. We propose an estimation scheme, catered to measurement data of this kind, to search for the exact maximum-likelihood-maximum-entropy estimator using semidefinite programming and a standard multi-dimensional function optimization routine. This scheme can be used to infer the expectation values of a set of entanglement witnesses that can be used to verify the entanglement of the unknown quantum state for composite systems. Next, we establish an alternative numerical scheme that is more computationally robust for the sole purpose of maximizing the likelihood and entropy.
\end{abstract}
\pacs{03.65.Ud, 03.65.Wj, 03.67.-a}
\submitto{\NJP}
\maketitle

\onehalfspacing
\section{Introduction}
Quantum state preparation is the first important step for any protocol that makes use of quantum resources. Examples of such protocols are quantum state teleportation and quantum key distribution which require entangled quantum states. In order to verify the integrity of the true quantum state $\rho_\mathrm{true}$ prepared by the source, one carries out quantum state tomography to characterize it. Measurements are performed on a collection of identical copies of quantum systems (electrons, photons, etc.) that are emitted from the source. Then, the quantum state of the source is inferred from the measurement data obtained from this collection. The measurements are generically described by a set of positive operators $\{\Pi_j\}$ that compose a \emph{probability operator measurement} (POM). Such a procedure of state inference is known as quantum state estimation.

When the measurement outcomes form an informationally complete set, they fully characterize the source and the measurement data obtained will contain maximal information about its state. To infer the unknown state from the data, one can search for a state estimator that maximizes the \emph{likelihood functional} that yields the probability of obtaining a particular sequence of measurement detections given a quantum state -- the maximum-likelihood (ML) estimator \cite{ml1,ml2,qstateest,rehacek1}. Yet, in tomography experiments performed on complex quantum systems with many degrees of freedom, it is not possible to implement such an informationally complete set of measurement outcomes. Therefore, some information about the source will be missing and its quantum state cannot be unambiguously determined. For instance, if a source produces a mode of light that is described by an infinite-dimensional statistical operator $\rho_\mathrm{true}$, then no matter how ingeniously a measurement scheme is designed to probe incoming photons prepared by this source, an infinite amount of information about the mode of light will always remain unknown. The ML estimator obtained from these informationally incomplete data is no longer unique and there will in general be infinitely many other ML estimators that are consistent with the data. These estimators form a convex set under the likelihood plateau.

In order to choose an estimator from the convex set for statistical prediction, we can consider the maximum-entropy principle advocated by E. T. Jaynes~\cite{jaynes1,jaynes2}. In doing so, one obtains a unique estimator that maximizes both the likelihood and the von Neumann entropy functional. Statistically, this estimator is least-biased for the informationally incomplete data. In \cite{ysteo1,ysteo2,ysteo3}, we developed and applied an algorithm, based on the steepest-ascent method, to approximately look for the maximum-likelihood-maximum-entropy (MLME) estimator. This algorithm involves a parameter that needs to be chosen just above a minimum threshold to obtain an estimator that is as close to the actual MLME estimator as possible. In general, this threshold depends on the true state $\rho_\mathrm{true}$. Therefore, one needs to run the algorithm a few times to estimate this threshold.

This paper is organized as follows. After a brief review of this steepest-ascent algorithm in Section~\ref{sec:br}, we introduce a numerical scheme that is based on completely different principles to directly search for the MLME estimator within the convex set in Section~\ref{sec:aasdp}. This scheme couples two separate optimization techniques --- semidefinite programming (SDP) and a derivative-free optimization method --- to maximize the entropy over linear combinations of a maximal set of linearly independent ML estimators that spans the ML convex set. It will be shown that, owing to the mechanisms of SDP, one can make use of this scheme to infer the expectation values of a set of entanglement witnesses to verify the presence of entanglement in the unknown quantum state for composite systems. Finally, in Section~\ref{sec:emlme}, we establish a more robust numerical scheme that systematically generates the maximal set of linearly independent ML estimators that defines the convex set without fail. Instead of SDP, this scheme utilizes a nonlinear optimization routine that finds the global maximum of a highly nonlinear functional that is used to generate this maximal set.

\section{Brief review}
\label{sec:br}
The likelihood functional $\log\mathcal{L}(\{n_j\};\rho)$ for a set of measurement data $\{n_j\}$ collected with a POM $\sum_j\Pi_j=1$ is given by\\
\begin{equation}
\mathcal{L}(\{n_j\};\rho)=\prod_jp_j^{n_j}\,,
\label{eq:like}
\end{equation}
where $n_j$ refers to the number of occurrences of the outcome $\Pi_j$ and $p_j=\tr\left\{\rho\Pi_j\right\}$. The corresponding frequencies are given by $f_j=n_j/N$. In ML state estimation, the concave log-likelihood functional $\log\mathcal{L}(\{n_j\};\rho)$ is maximized to obtain the ML estimator $\hat\rho_\mathrm{ML}$ \footnote{The symbol ``\,$\hat{\,\,\,}$\,'' is used to denote all estimators.} for the given set of data. If the number of linearly independent outcomes is $D^2$, with $D$ being the dimension of the Hilbert space, the data is informationally complete and the estimator $\hat\rho_\mathrm{ML}$ is unique. In the case where this number is less than $D^2$, the data is informationally incomplete and there exists now a continuous set of $\hat\rho_\mathrm{ML}$ that yield the same maximal likelihood. This set is convex since the likelihood functional is concave in $\rho$.

To choose the estimator $\hat\rho_\mathrm{MLME}$ that has the highest entropy out of this convex set, we can consider the Lagrange functional\\
\begin{equation}
\mathcal{I}(\lambda;\rho)=\lambda\left(S(\rho)-S_\mathrm{max}\right)+\frac{1}{N}\log\mathcal{L}(\{n_j\};\rho)\,,
\label{eq:newinfo}
\end{equation}
where $\lambda$ is the Lagrange multiplier corresponding to the constraint of maximal entropy $S(\hat\rho_\mathrm{MLME})=-\tr\{\hat\rho_\mathrm{MLME}\log\hat\rho_\mathrm{MLME}\}=S_\mathrm{max}$. In doing so, we maximize the two functionals $S(\rho)$ and $\log(\mathcal{L}(\{n_j\};\rho))$ simultaneously. We define $\hat\rho_{\mathrm{I},\lambda}$ to be the estimator that maximizes $\mathcal{I}(\lambda;\rho)$.

If $\lambda=0$, $\mathcal{I}(\lambda;\rho)=\log\mathcal{L}(\{n_j\};\rho)/N$ and maximizing this Lagrange functional is just the procedure of ML. Since the data is informationally incomplete, there exists a convex plateau structure for the log-likelihood functional and maximizing $\mathcal{I}(\lambda;\rho)$ yields a convex set of estimators. For large $\lambda$ values, the term $\lambda S(\rho)$ dominates, so that the resulting estimator $\hat\rho_{\mathrm{I},\lambda\rightarrow\infty}=1/D$. When $\lambda$ takes on a very small positive value \cite{ysteo1}, the contribution from $\lambda S(\rho)$ becomes relatively much smaller than that of $\log(\mathcal{L}(\{n_j\};\rho))/N$, and any variation of the von Neumann entropy functional is only detectable over the state space region that coincides with the likelihood plateau. In other words, maximizing $\mathcal{I}(0<\lambda\rightarrow0;\rho)$ is equivalent to maximizing the entropy over the plateau. Therefore, $\hat\rho_{\mathrm{I},0<\lambda\rightarrow0}=\hat\rho_\mathrm{MLME}$. The iterative algorithm that is based on the steepest-ascent method is described in \cite{ysteo1}-\cite{ysteo3}.

In practice, there is a limit to how small $\lambda$ can be. In particular, when $\lambda$ is smaller than some numerical threshold $\lambda_\mathrm{thres}>0$, the gradient of $\lambda S(\rho)$ is no longer visible. In this case, the algorithm treats $\mathcal{I}(\lambda<\lambda_\mathrm{thres};\rho)$ as $\mathcal{I}(\lambda=0;\rho)$ and performs ML estimation. Hence, the optimal parameter $\lambda$ is to be slightly above $\lambda_\mathrm{thres}$ so that $\hat\rho_{\mathrm{I},\lambda\gtrsim\lambda_\mathrm{thres}}\approx\hat\rho_\mathrm{MLME}$. This inevitably introduces a small bias to the estimator. Determining the value of $\lambda_\mathrm{thres}$ analytically is quite complicated because $\lambda_\mathrm{thres}$ is actually a function of the true state and the POM, that is $\lambda_\mathrm{thres}=\lambda_\mathrm{thres}(\{\Pi_j\};\rho_\mathrm{true})$. Since $\rho_\mathrm{true}$ is unknown to us, one needs to estimate $\lambda_\mathrm{thres}$ through repeated runs of the algorithm. In the next section, we will introduce an alternative scheme to search for the exact MLME estimator.

\section{Algorithm for entanglement and state verification}
\label{sec:aasdp}
To obtain the unique MLME estimator that has the highest entropy, a search has to be performed within the ML convex set. In this section, we introduce a numerical procedure to estimate $\hat\rho_\mathrm{MLME}$ in two steps. The first step is to obtain a collection of boundary states of the convex set for the measurement data obtained. In the next step, the operator $\hat\rho_\mathrm{MLME}$ can be estimated using these boundary states with a standard function optimization routine.

To carry out the first step, we need to identify the boundary of the ML convex set. Especially for large dimensions, the boundary of the convex set has an extremely complicated geometry that is too difficult to be analytically determined. Instead, we investigate its boundary by numerical means. We begin with the fact that the real functional $f(H;\rho)=\tr\{\rho H\}$, where the operator $H=H^\dagger$, is a linear functional of $\rho$. Therefore, if we try to maximize (or minimize) $f(H;\rho)$ over some subspace of $\rho$ that has a well defined boundary, the maxima (or minima) of this linear functional are always on the boundary of this subspace. We can thus generate boundary states by maximizing or minimizing $f(H;\rho)$, for a given Hermitian operator $H$, over the ML convex set. This problem is equivalent to the following optimization task:\\
\begin{center}
\colorbox{white}{\begin{minipage}[c]{10cm}
Maximize or minimize $f(H;\rho)=\tr\{\rho H\}$, subjected to the following constraints:
\begin{itemize}
  \item $\rho\geq0$,
  \item $\tr\{\rho\}=1$,
  \item $\tr\{\rho\Pi_j\}=\tr\{\hat\rho_\mathrm{ML}\Pi_j\}$ for all $j$.\\
\end{itemize}
\end{minipage}}
\end{center}
This is a standard linear optimization problem, with linear and positivity constraints, that can be solved with the help of SDP \cite{sdp}.

At this point, we need to find out the minimum number of boundary states that is required to search for $\hat\rho_\mathrm{MLME}$. To do this, we represent a generic ML estimator $\hat\rho_\mathrm{ML}$ by a set of $D^2$ linearly independent\footnote{If $L$ Hermitian operators $A_j$ are all linearly independent, the rank of the matrix $M$ with elements $M_{jk}=\tr\{A_jA_k\}$ is $L$.} trace-orthonormal Hermitian basis operators $\Gamma_j=\Gamma^\dagger_j$ satisfying the condition $\tr\{\Gamma_j\Gamma_k\}=\delta_{jk}$. With these basis operators, we can separate
\begin{equation}
\hat\rho_\mathrm{ML}=\underbrace{\sum^{D_\mathrm{meas}}_{j=1}a_j\,\Gamma^\mathrm{\,meas}_j}_{\displaystyle =\tilde\rho_\mathrm{meas}}+\underbrace{\sum^{D_\mathrm{unmeas}}_{j=1}b_j\,\Gamma^\mathrm{\,unmeas}_j}_{\displaystyle =\tilde\rho_\mathrm{unmeas}}
\label{eq:basisopexp}
\end{equation}
into the part in the measurement subspace of dimension $D_\mathrm{meas}$, and the rest of the state space that constitutes the unmeasured parameters (the unmeasured subspace) of dimension $D_\mathrm{unmeas}$, where $D_\mathrm{meas}+D_\mathrm{unmeas}=D^2$, $a_j=\tr\{\hat\rho_\mathrm{ML}\,\Gamma^\mathrm{\,meas}_j\}$ and $b_j=\tr\{\hat\rho_\mathrm{ML}\,\Gamma^\mathrm{\,unmeas}_j\}$. The generation of the basis operators $\Gamma^\mathrm{\,meas}_j$ that span the measurement subspace can be done by the Gram-Schmidt orthonormalization procedure on the $\Pi_j$s, with all conventional inner products for vectors replaced by trace inner products for operators. In other words, the number of linearly independent POM outcomes $\Pi_j$ is $D_\mathrm{meas}$. To generate the rest of the basis operators that span the unmeasured subspace, we continue the Gram-Schmidt procedure using randomly generated positive operators instead of the $\Pi_j$s.

From \eref{eq:basisopexp}, it can be deduced that the maximum dimension of the ML convex set is $D_\mathrm{unmeas}$. To show this, we note that \emph{every} ML estimator contains the same operator $\tilde\rho_\mathrm{meas}$ since the probabilities of $p_j=\tr\{\tilde\rho_\mathrm{meas}\Pi_j\}$ are fixed and $\tr\{\tilde\rho_\mathrm{unmeas}\Pi_j\}=0$ for all $\Pi_j$ by definition of the trace-orthonormal basis operators. This implies that the only difference between any two ML estimators in the convex set is $\tilde\rho_\mathrm{unmeas}$. As the operators $\tilde\rho_\mathrm{meas}$ and $\tilde\rho_\mathrm{unmeas}$ are linearly independent, it follows that any ML estimator can always be expressed as a linear combination of the unique $\tilde\rho_\mathrm{meas}$ and the $D_\mathrm{unmeas}$ linearly independent basis operators that define $\tilde\rho_\mathrm{unmeas}$. This means that a set of $D_\mathrm{unmeas}+1$ linearly independent boundary states is enough to look for the MLME estimator. As the operator $\tilde\rho_\mathrm{meas}$ is fixed by the measurement operators, the maximal number of free parameters that span the convex set is $D_\mathrm{unmeas}$. In some cases, however, the dimension of the ML convex set is lower due to the positivity constraint imposed on the ML estimators. In the extreme case, the convex set is restricted to a single point in state space. In these situations, we do not know its actual dimension and repeated generation of boundary states is necessary to estimate the maximum number of linearly independent boundary states. We remind the reader that with enough linearly independent states, the exact MLME estimator can be obtained up to numerical precision. In Section~\ref{sec:emlme}, a different numerical procedure will be introduced to generate the maximal set of linearly independent ML estimators that spans the ML convex set without requiring any knowledge of the convex set.

Apart from serving as a routine to numerically compute the boundary states of the convex set, SDP provides an additional useful function. For composite quantum systems, if one selects the Hermitian operators $H$ to be entanglement witnesses $W$, one can obtain information about the presence of entanglement in the unknown true state even with informationally incomplete data. These witnesses have the properties that $\tr\{\rho_\mathrm{sep}W\}\geq0$ for all separable states $\rho_\mathrm{sep}$ and $\tr\{\rho_\mathrm{ent}W\}<0$ for at least one entangled state $\rho_\mathrm{ent}$. The SDP routine, described above, thus looks for the maximum (or minimum) value of $f(W;\rho)$ for any chosen witness operator $W$ over the space of positive $\rho$s. This way, we can in fact infer a set of maximum (or minimum) witness expectation values over the ML convex set from this optimization procedure. The set of inferred maximum witness expectation values is particularly informative, for if the maximum value of $f(W;\rho)$ for at least one of the randomly generated operators $W$ is negative, we can immediately conclude that the true state is entangled since this state must lie within the ML convex set that results from the incomplete data. For practical computation, we can choose the witness operators $W$ to be of the decomposable form $W=Q^{\textsc{t}_j}$, where $Q$ is a positive operator with no product kets in its range and the symbol ``\,$\textsc{t}_j$'' denotes a partial transposition with respect to the $j$th subsystem. For bipartite systems, these operators are optimal witnesses \cite{wit1,wit2}, that is no other witnesses can detect all entangled states that are detected by this witness, as well as other entangled states. For multipartite systems, these operators still serve as entanglement witnesses since $\tr\left\{\rho_\mathrm{sep}\,Q^{\textsc{t}_2}\right\}=\tr\left\{\rho^{\textsc{t}_2}_\mathrm{sep}\,Q\right\}>0$, although they are no longer optimal in general. To obtain a random set of boundary states of the ML convex set, random statistical operators $Q$ are generated using the relation\\
\begin{equation}
Q=\frac{X^\dagger X}{\tr\left\{X^\dagger X\right\}}\,,
\label{eq:wishart}
\end{equation}
where $X$ is a random operator which, when expressed in the computational basis, has complex matrix elements that are distributed according to the standard normal distribution of zero mean and unit variance.

The second step involves an optimization procedure to maximize the entropy $S(\rho)$ using the generated set of $M_0\leq D_\mathrm{unmeas}+1$ linearly independent boundary states $\left\{\rho^{(j)}_\mathrm{bd}\right\}^{M_0}_{j=1}$. For the purpose of entanglement detection, these boundary states are obtained by maximizing the linear functionals $f(W;\rho)$ of a set of witness operators $W$ over the ML convex set according to the recipe described above. We start by writing a generic ML estimator as a linear combination of $\rho^{(j)}_\mathrm{bd}$ inasmuch as\\
\begin{equation}
\hat\rho_\mathrm{ML}\left(\{t_j\}\right)=\sum^{M_0}_{j=1}t_j\,\hat\rho^{(j)}_\mathrm{bd}\,,
\label{eq:lincomb}
\end{equation}
where the $t_j$s are normalized coefficients, such that $\sum_jt_j=1$, that are in general real such that $\hat\rho_\mathrm{ML}\left(\{t_j\}\right)\geq0$. The task now is to look for the values of $t_j=t^{\,\mathrm{max}}_j$ for which the function $S\left(\{t^{\,\mathrm{max}}_j\}\right)=S\left(\hat\rho_\mathrm{ML}\left(\{t^{\,\mathrm{max}}_j\}\right)\right)$ is maximum over all real normalized coefficients. The unconstrained optimization of this $M_0$-dimensional function with respect to $t_j$ can be performed with any efficient multi-dimensional unconstrained optimization routine that is included in the standard libraries of commercialized mathematical softwares. In MATLAB, for instance, the function \texttt{fminsearch} does the job using the Nelder-Mead simplex method (NMS). When $M_0$ is large, it is suggested in \cite{anms} that an adaptive version of the Nelder-Mead simplex method (ANMS) may be more advantageous in terms of shorter computation time. We take the resulting operator $\hat\rho_\mathrm{SDP}\equiv\hat\rho_\mathrm{ML}\left(\{t^{\,\mathrm{max}}_j\}\right)$ as the SDP~MLME estimator. There is, however, a caveat to this optimization. Since the positivity of $\hat\rho_\mathrm{ML}\left(\{t_j\}\,\right)$ is no longer guaranteed over the entire space of real normalized vectors, the entropy\\
\begin{equation}
S(\{\nu_j\})=-\sum^D_{j=1}\nu_j\log\nu_j\,,
\label{eq:origent}
\end{equation}
expressed in terms of the eigenvalues $\nu_j$ of the statistical operator $\hat\rho_\mathrm{ML}\left(\{t_j\}\right)=\sum_j\ket{\nu_j}\nu_j\bra{\nu_j}$ that is represented by its eigenbasis $\{\ket{\nu_j}\}$, can take complex values. In order to restrict the optimization to yield only positive SDP~MLME estimators, one can replace the entropy function in \eref{eq:origent} with the conditional function\\
\begin{equation}
S_\mathrm{cond}(\{\nu_j\}) =
\cases{-\sum^D_{j=1}\nu_j\log\nu_j & for\,\,$\hat\rho_\mathrm{ML}\left(\{t_j\}\right)\geq0$\,,\\
S_0<0 & otherwise\,,}
\label{eq:condent}
\end{equation}
which effectively restricts the original search region to the admissible state space. To check if the evaluated operator $\hat\rho_\mathrm{ML}\left(\{t_j\}\right)$ is positive, a highly efficient way is to determine whether or not it admits a Cholesky decomposition.

For full-rank SDP~MLME estimators $\hat\rho_\mathrm{SDP}$, there is a simple way to check that $\hat\rho_\mathrm{SDP}$ is indeed the MLME estimator. We recall that the form of the MLME estimator is given by
\begin{equation}
\hat\rho_\mathrm{MLME}=\frac{\rme^{\sum_j\lambda_j\Pi_j}}{\tr\left\{\rme^{\sum_k\lambda_k\Pi_k}\right\}}\,,
\label{eq:mlmeexpr}
\end{equation}
where each $\lambda_j$ is a Lagrange multiplier for the constraint $\tr\{\hat\rho_\mathrm{MLME}\,\Pi_j\}=\tr\{\hat\rho_\mathrm{ML}\,\Pi_j\}$ for any ML estimator $\hat\rho_\mathrm{ML}$ in the convex set. If $\hat\rho_\mathrm{MLME}$ is full-rank, it follows from \eref{eq:mlmeexpr} that the operator $\log\hat\rho_\mathrm{MLME}$ is a linear combination of only the POM outcomes, that is $\log\hat\rho_\mathrm{MLME}$ resides in the measurement subspace. This is equivalent to the set of conditions $\tr\{\Gamma^\mathrm{\,unmeas}_j\,\log\hat\rho_\mathrm{SDP}\}=0$ for all $\Gamma^\mathrm{\,unmeas}_j$s. Defining the variables $c_j=\tr\{\Gamma^\mathrm{\,unmeas}_j\,\log\hat\rho_\mathrm{SDP}\}$, the quantity $\gamma\equiv\sqrt{\sum_jc_j^2}$ can be used to determine if $\hat\rho_\mathrm{SDP}$ is close to the actual MLME estimator up to some numerical precision.

\begin{figure}[!h]
\centering
  \subfloat[]{\label{fig:mepart}\includegraphics[width=0.55\columnwidth]{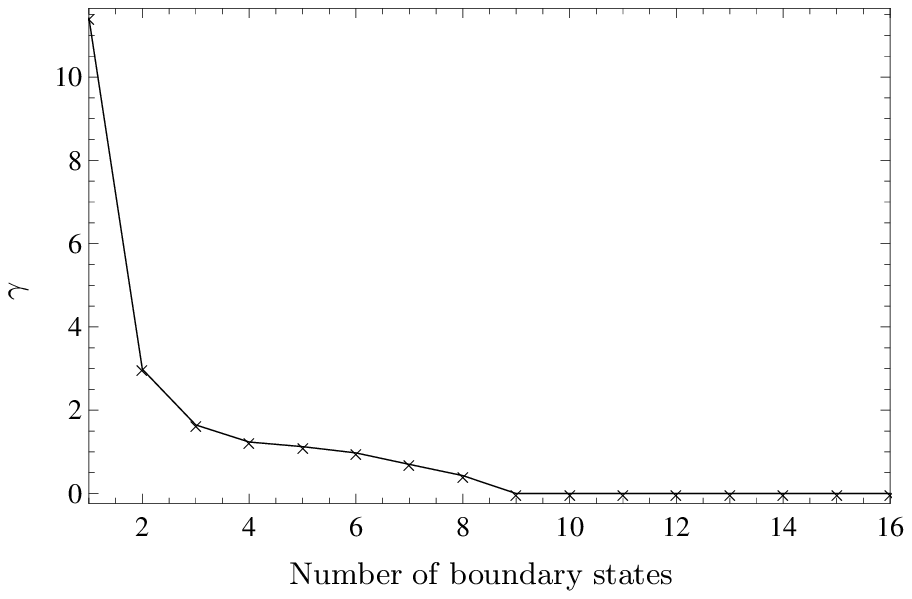}}\\
  \subfloat[]{\label{fig:ent_det_ratio}\includegraphics[width=0.55\columnwidth]{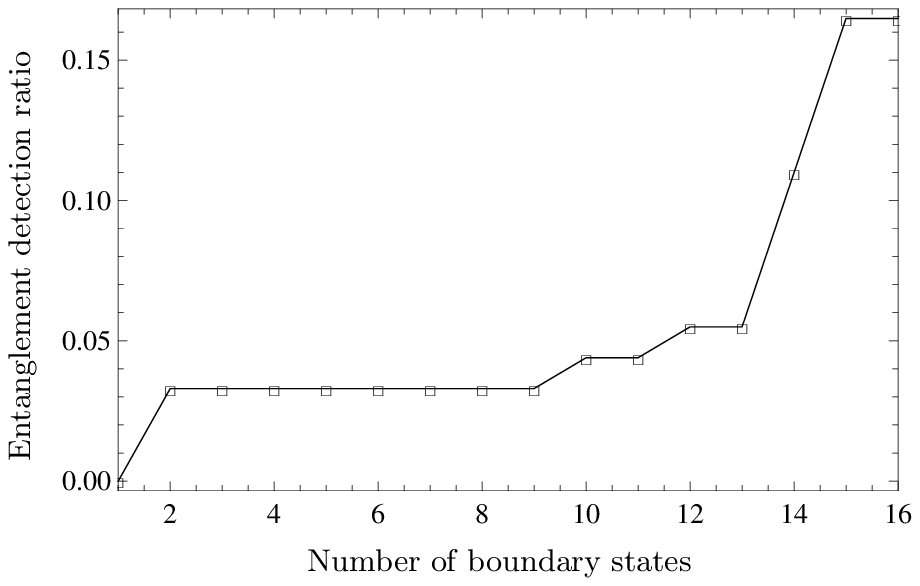}}\\
  \subfloat[]{\label{fig:mlme_dist}\includegraphics[width=0.55\columnwidth]{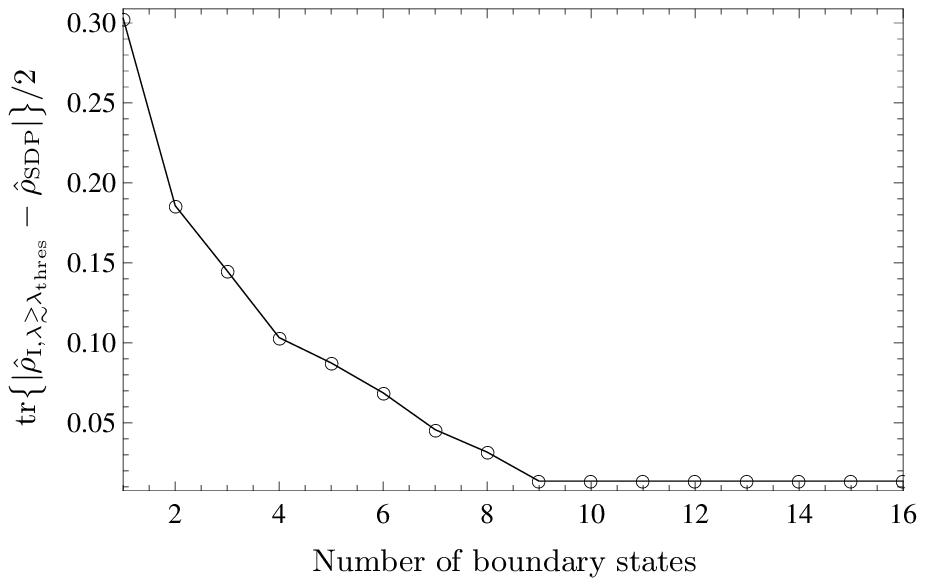}}
    \caption{Three plots showing various aspects of SDP~MLME conducted on 100 random two-qubit entangled pure states as functions of the number of linearly independent ML boundary states used, with plots (a) and (c) averaged over all these pure states. A randomly-generated eight-outcome informationally incomplete POM is used throughout the simulations. In this case, every corresponding ML convex set is specified by nine linearly independent ML estimators. Decomposable optimal witness operators of the form $Q^{\textsc{t}_2}$, where the operators $Q$ are entangled pure states, are used to generate the boundary states with SDP.}
  \label{fig:two_qubit}
\end{figure}

In summary, both the MLME estimator and information about the entanglement of $\rho_\mathrm{true}$ can be obtained with informationally incomplete data using the following algorithm, which we coin as SDP~MLME:\\
\begin{center}
\colorbox{white}{\begin{minipage}[c]{10cm}
\textbf{SDP~MLME}
\begin{enumerate}
\item \textbf{ML} --- Obtain the ML probabilities for the POM used to collect the measurement data with the ML algorithm.
\item \textbf{SDP} --- Perform SDP, using \eref{eq:wishart}, to obtain the maximal set of $M_0\leq D_\mathrm{unmeas}+1$ linearly independent boundary states that define the ML convex set using witness operators. Inspect the list of maximum expectation values of the witness operators and conclude that the unknown quantum state is entangled if any one of them is negative.
\item \textbf{Entropy maximization} --- Carry out function maximization with NMS or ANMS on the entropy function $S\left(\{t_j\}\right)$ with respect to all real normalized coefficients $t_j$ using \eref{eq:lincomb} and \eref{eq:condent} to obtain $\hat\rho_\mathrm{SDP}$.\\
\end{enumerate}
\end{minipage}}
\end{center}
Figure \ref{fig:two_qubit} summarizes the results.
In Figure~(\ref{fig:mepart}), the convergence of $\gamma$ is consistent with the fact that, in principle, the maximal number of nine linearly independent boundary states is enough to express the MLME estimator. The entanglement detection ratio in Figure~\ref{fig:ent_det_ratio} is computed as the ratio of the number of detected pure states to the total number of generated states for every number of boundary states used. Ideally, this ratio eventually goes to one if enough witness operators are generated to detect all the random pure states since there are no positive partial-transpose entangled states in this case \cite{wit1,wit2}. For benchmarking, Figure~\ref{fig:mlme_dist} is generated to confirm the consistency of SDP~MLME with the MLME algorithm described in Section~\ref{sec:br}. There exists an average bias in \ref{fig:mlme_dist} that arises from a fixed $\lambda_\mathrm{thres}=10^{-5}$ for all the pure states.

\section{Robust algorithm for incomplete state estimation}
\label{sec:emlme}
Despite the usefulness of SDP in entanglement verification and boundary states generation as discussed in Section \ref{sec:aasdp}, the speed of the SDP routine strongly depends on the dimension of the Hilbert space and the total number of linear constraints imposed by the measurement data. When the total number of POM outcomes is large, there will generally be a considerable slowdown of the SDP routine as the search accounts for a large set of linear constraints in addition to the positivity constraint. Another feature of the SDP routine is that the sequence of boundary states that are generated from random Hermitian operators are not guaranteed to be linearly independent of one another. This means that typically, one would need to generate a large set of ML estimators that contains the maximal number of linearly independent estimators that span the ML convex set. For convex sets of large dimensions, this approach can be time-consuming. In this section, we propose a different search routine, in place of SDP, to directly look for linearly independent ML estimators within the ML convex set in a deterministic way. With this routine, we establish a feasible algorithm to look for the exact MLME estimator.

To begin, we recall that a given set of $M_0$ linearly independent ML estimators, as in \eref{eq:lincomb}, implies the existence of a full-rank $M_0\times M_0$ positive Gram matrix $M_\textsc{g}$ with elements given by\\
\begin{equation}
\left(M_\textsc{g}\right)_{jk}=\frac{\tr\left\{\hat\rho^{(j)}_\mathrm{ML}\,\hat\rho^{(k)}_\mathrm{ML}\right\}}{\sqrt{\tr\left\{\left(\hat\rho^{(j)}_\mathrm{ML}\right)^2\right\}\tr\left\{\left(\hat\rho^{(k)}_\mathrm{ML}\right)^2\right\}}}\,.
\end{equation}
This hints a straightforward strategy to cumulatively obtain the maximal set of linearly independent ML estimators that spans the entire ML convex set: Starting with a single ML estimator, the next estimator containing the same $\tilde\rho_\mathrm{meas}$ should be chosen such that the smallest eigenvalue $\sigma_\mathrm{min}\left(M_\textsc{g}\right)$ of the Gram matrix $M_\textsc{g}$ for these two estimators is maximized, and so forth, with the maximization performed over positive estimators $\hat\rho^{(j)}_\mathrm{ML}\geq0$.

In general, $\sigma_\mathrm{min}(M_\textsc{g})$ is a nonlinear functional of $M_\textsc{g}$ that has multiple local stationary points. This functional is also not differentiable and has undefined gradients at the boundary of the state space. To search for its global maximum, an appropriate numerical method to use is a nonlinear optimization algorithm that invokes pattern searches \cite{pattern1} and can cope with functionals that have ill-defined gradients. The solver for this algorithm, \texttt{patternsearch}, is readily available in MATLAB. The positivity constraint $\hat\rho^{(j)}_\mathrm{ML}\geq0$ is incorporated into the optimization algorithm using the augmented Lagrangian method with Cholesky decomposition. Another versatile feature of the proposed routine is that it can be set to terminate when the maximal number of linearly independent ML estimators is generated, such that the next ML estimator always yields a zero eigenvalue for $M_\textsc{g}$. We can, therefore, deterministically obtain the maximal set of linearly independent ML estimators that spans the ML convex set in this manner, in contrast with the SDP routine in SDP~MLME. In this sense, the routine is operationally robust even without any prior information about the convex set. To ensure that the search is numerically stable, it is favorable to start the pattern search algorithm from a highly-mixed ML estimator. This is obtained by performing SDP as a couple of times and defining the starting estimator as an equal mixture of the resulting ML boundary estimators and the fairly mixed ML estimator obtained from the ML algorithm starting from the maximally-mixed state in computing the ML probabilities. Using this maximal set, the MLME estimator can be directly obtained by maximizing the entropy function in \eref{eq:condent} over all linear combinations of the linearly independent ML estimators in the set. These lead to the following pattern search MLME algorithm (PS~MLME) for incomplete quantum state estimation:\\
\begin{center}
\colorbox{white}{\begin{minipage}[c]{10cm}
\textbf{PS~MLME}
\begin{enumerate}
\item \textbf{ML} --- Obtain the ML probabilities, as well as the corresponding ML estimator that is fairly mixed, for the POM used to collect the measurement data with the ML algorithm starting from the maximally-mixed state.
\item Carry out SDP a couple of times to obtain a few ML boundary estimators and check if the ML estimators are close to each other. If the average pairwise norm is smaller than a certain threshold, this means that the ML convex set can be approximated to be a single point and the ML estimator is taken to be the unique estimator. Otherwise, proceed to the next step.
\item \textbf{Definition of the ML convex set} --- Generate the maximal set of linearly independent ML estimators that defines the ML convex set by maximizing $\sigma_\mathrm{min}(M_\textsc{g})$ via the augmented Lagrangian pattern search algorithm.
\item \textbf{Entropy maximization} --- Carry out function maximization with NMS or ANMS on the entropy function $S(\{t_j\})$ with respect to all real normalized coefficients $t_j$ using \eref{eq:lincomb} and \eref{eq:condent} to obtain the MLME estimator.\\
\end{enumerate}
\end{minipage}}
\end{center}

\begin{figure}[!h]
\centering
  \subfloat[]{\label{fig:cputime}\includegraphics[width=0.55\columnwidth]{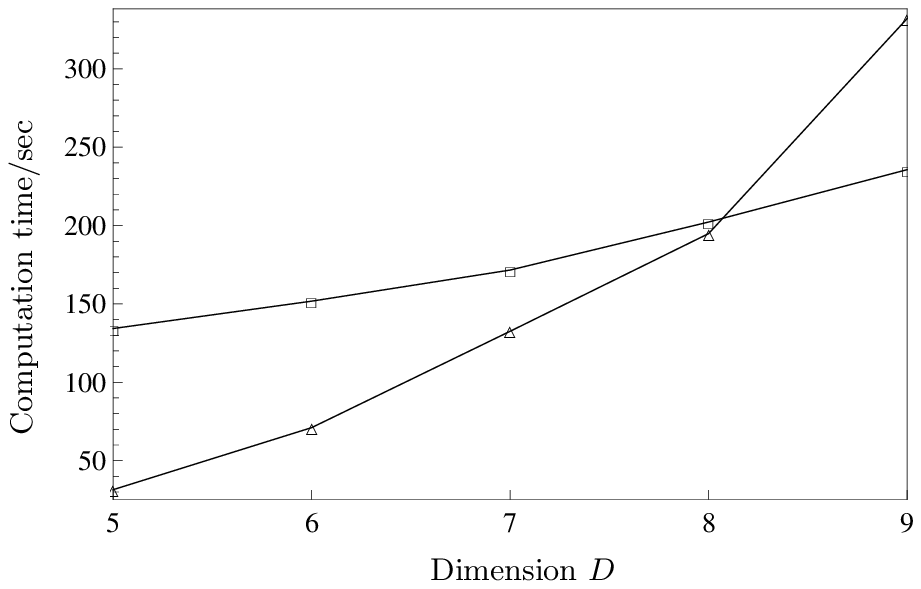}}\\
  \subfloat[]{\label{fig:accuracy}\includegraphics[width=0.55\columnwidth]{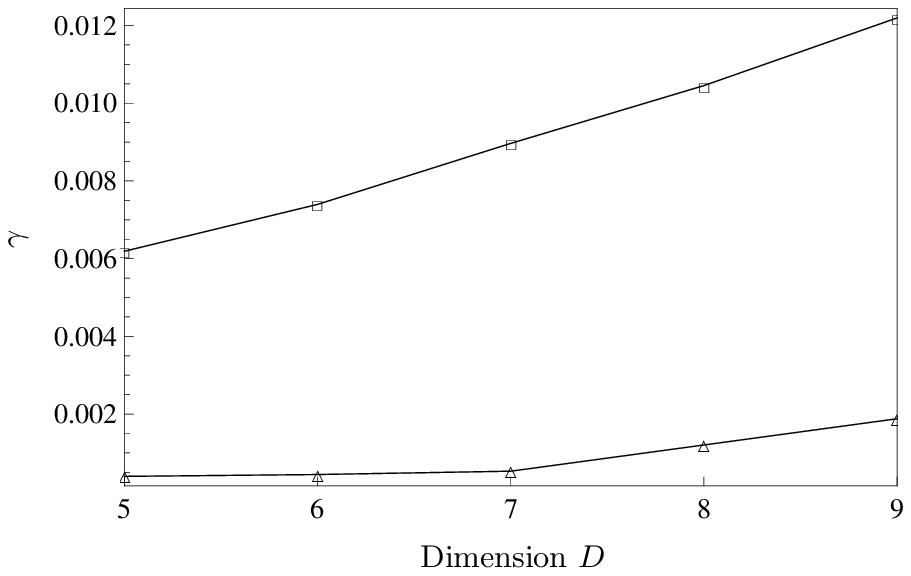}}\\
    \caption{Plots showing the average computation time for PS~MLME ($\triangle$) and steepest-ascent MLME ($\square$) with $\lambda_\mathrm{thres}=10^{-5}$, as well as their respective average accuracies of the estimators for various Hilbert space dimensions $D$. The random POM for each dimension contains $\lfloor D^2/2\rfloor$ outcomes that are linearly independent. Each data point is averaged over 50 randomly generated full-rank true states of every dimension. The computation time is for MATLAB running on an Intel i7 2.67 GHz Quad Core personal computer, where all function tolerances are set to $10^{-8}$.}
  \label{fig:cvx_sa}
\end{figure}

The performance of PS~MLME depends not only on the dimension $D_\mathrm{unmeas}$ of the unmeasured subspace, but also on the complexity of the functional $\sigma_\mathrm{min}\left(M_\textsc{g}\right)$. More generally, as the dimension of the Hilbert space, or that of the unmeasured subspace, increases, the number of local maxima of $\sigma_\mathrm{min}\left(M_\textsc{g}\right)$ increases. This translates to a longer computation time to locate the global maximum in the search for linearly independent ML estimators. Thus, for very large dimensions, PS~MLME becomes inefficient and the approximate MLME algorithm, which is based on steepest ascent \cite{ysteo1}, is a more practical substitute. On the other hand, PS~MLME consistently gives more accurate MLME estimators as compared to the steepest-ascent algorithm, which yields biased results for large dimensions because of the finite $\lambda$ parameter. Hence, there is a tradeoff between computation time and the accuracy of the MLME estimators. Figure~\ref{fig:cvx_sa} compares the performances of the PS~MLME and the steepest-ascent MLME algorithms for varying Hilbert space dimensions.
Figure~\ref{fig:accuracy} shows the consistently more accurate results obtained with PS~MLME as compared to the steepest-ascent algorithm, in exchange for its longer computation time, illustrated in Figure~\ref{fig:cputime}, for higher dimensions. The simulations show that for the moderately large dimensions considered in Figure~\ref{fig:cvx_sa}, much more accurate MLME estimators can be obtained using PS~MLME with longer computation time that is of the same order as that with steepest ascent. To reduce the computation time of PS~MLME, one can consider an approximate maximization of $\sigma_\mathrm{min}\left(M_\textsc{g}\right)$ as long as the resulting value is sufficiently large. Throughout the simulations, the duration of searching for each optimal ML estimator is restricted to five seconds. For even larger dimensions, PS~MLME can be computationally demanding even when approximate maximization is carried out, and the steepest-ascent algorithm turns out to be a more realistic option. Efforts to further improve the performance of the PS~MLME scheme are in the works. Nevertheless, we hope that the current work can serve as a stepping stone that helps to spur interesting discussions and novel contributions in related fields on numerical optimization over convex sets of positive operators.

\section{Conclusion}
We have introduced a scheme to look for the unique estimator that maximizes the likelihood and entropy for informationally incomplete measurement data. This involves two main procedures: a generation of linearly independent boundary maximum-likelihood estimators that spans the convex set with semidefinite programming, and an entropy maximization with these estimators using a standard function optimization routine. Furthermore, for composite quantum systems, one can apply this scheme to infer the expectation values of a set of entanglement witnesses. This information allows us to verify the entanglement of any quantum state with informationally incomplete data. However, semidefinite programming does not offer a definite control over the generation of maximum-likelihood estimators. This motivated us to develop an alternative scheme that is more operationally robust than the former one to search for the maximum-likelihood-maximum-entropy estimator. This latter scheme makes use of the pattern search optimization algorithm that is suitable for maximizing a nonlinear function required to deterministically generate the maximal set of estimators that defines the convex set. With numerical simulations, we showed that the latter scheme gives much more reliable results than the MLME algorithm discussed in Section~\ref{sec:br} at the expense of slightly longer computation time when the dimension of the reconstruction Hilbert space, or the unmeasured subspace, is moderately large.

\ack
This work is supported by the NUS Graduate School for Integrative Sciences and Engineering and the Centre for Quantum Technologies, which is a Research Centre of Excellence funded by Ministry of Education and National Research Foundation of Singapore, as well as the Technology Agency of the Czech Republic, Project No. TE01020229 (Center
of Digital Optics), and the Czech Ministry of Industry and Trade, Project No. FR-TI1/364.


\section*{References}
\begin{thebibliography}{99}
\bibitem{ml1}
Fisher R A 1922 {\it Phil. Trans. R. Soc. London} {\bf222} 309
\bibitem{ml2}
Helstr{\o}m C W 1976 {\it Quantum Detection and Estimation Theory} (New York: Academic Press)
\bibitem{qstateest}
{\v R}eh{\'a}{\v c}ek J and Paris M 2004 {\it Lecture {Notes} in {Physics} -- {Quantum} {State} {Estimation}} vol~649 (Berlin Heidelberg: Springer)
\bibitem{rehacek1}
{\v R}eh{\'a}{\v c}ek J, Hradil Z, Knill E and Lvovsky A I 2007 {\it\PR~A} {\bf75} 042108
\bibitem{jaynes1}
Jaynes E T 1957 {\it\PR} {\bf106} 620
\bibitem{jaynes2}
Jaynes E T 1957 {\it\PR} {\bf108} 171
\bibitem{ysteo1}
Teo Y S, Zhu H, Englert B-G, {\v R}eh{\'a}{\v c}ek J and Hradil Z 2011 {\it\PRL} {\bf10} 020404
\bibitem{ysteo2}
Teo Y S, Englert B-G, {\v R}eh{\'a}{\v c}ek J and Hradil Z 2011 {\it\PR~A} {\bf84} 062125
\bibitem{ysteo3}
Teo Y S, Stoklasa B, Englert B-G, {\v R}eh{\'a}{\v c}ek J and Hradil Z 2012 {\it\PR~A} {\bf85} 042317
\bibitem{sdp}
Vandenberghe L and Boyd S 1996 {\it SIAM~review} {\bf38} 49
\bibitem{wit1}
Zhu H, Teo Y S and Englert B-G 2010 {\it\PR~A} {\bf81} 052339
\bibitem{wit2}
Lewenstein M, Kraus B, Cirac J I and Horodecki P 2000 {\it\PR~A} {\bf62} 052310
\bibitem{anms}
Gao F and Han L 2010 {\it Comput.~Optim.~Appl.} {\bf51} 259
\bibitem{pattern1}
Audet C and Dennis Jr J E 2003 {\it SIAM J. Opt.} {\bf13} 889
\end{thebibliography}
\end{document}